\input epsf
\input harvmac
\noblackbox


\def\narrowplus{\kern -.04truein + \kern -.03truein}
\def\narrowminus{- \kern -.04truein}
\def\narrowminussub{\kern -.02truein - \kern -.01truein}

\def\frac#1#2{{#1\over #2}}

\def\lappeq{\mathrel{\rlap{\raise.5ex\hbox{$<$}}
            {\lower.5ex\hbox{$\sim$}}}}

%
%
\def\eqnn#1{\xdef #1{(\secsym\the\meqno)}\writedef{#1\leftbracket#1}%
\global\advance\meqno by1\wrlabeL#1}
\def\eqna#1{\xdef #1##1{\hbox{$(\secsym\the\meqno##1)$}}
\writedef{#1\numbersign1\leftbracket#1{\numbersign1}}%
\global\advance\meqno by1\wrlabeL{#1$\{\}$}}
\def\eqn#1#2{\xdef #1{(\secsym\the\meqno)}\writedef{#1\leftbracket#1}%
\global\advance\meqno by1$$#2\eqno#1\eqlabeL#1$$}

\newcount\figno
\figno=0
\def\fig#1#2#3{
\par\begingroup\parindent=0pt\leftskip=1cm\rightskip=1cm\parindent=0pt
\baselineskip=11pt
\global\advance\figno by 1
\midinsert
\vskip -9ex
\epsfxsize=#3
\centerline{\epsfbox{#2}}
\vskip 12pt
{\bf Fig.\ \the\figno: } #1\par
\endinsert\endgroup\par
}
\def\figlabel#1{\xdef#1{\the\figno}}
\def\encadremath#1{\vbox{\hrule\hbox{\vrule\kern8pt\vbox{\kern8pt
\hbox{$\displaystyle #1$}\kern8pt}
\kern8pt\vrule}\hrule}}
\lref\aadd{N. Arkani-Hamed, S. Dimopoulos and G. Dvali, {\it The hierarchy
problem and new dimensions at a millimeter}, hep-ph/9803315, Phys. Lett 
{\bf B429}(1998) 263; {\it Phenomenology, astrophysics and cosmology of
theories with sub-millimeter dimensions and TeV scale
quantum gravity}, hep-ph/9807344, Phys. Rev. {\bf D59} (1999) 086004;
\hfill\break

I. Antoniadis, N. Arkani-Hamed, S. Dimopoulos and G. Dvali, {\it New
dimensions at a millimeter to a fermi and superstrings at a TeV},
hep-ph/9804398, Phys. Lett {\bf B436} (1998) 257.}

\lref\rstwo{L. Randall and R. Sundrum, {\it An Alternative to 
            compactification}, hep-th/9906064.}

\lref\gw{W. D. Goldberger and Mark B. Wise, {\it Bulk Fields in the 
Randall-Sundrum Compactification Scenario}, hep-ph/9907218; 
{\it Modulus stabilization with bulk fields}, hep-ph/9907447.}

\lref\kmprs{I.I. Kogan, S. Mouslopoulos, A. Papazoglou, G.G. Ross
and J. Santiago, {\it A Three three-brane Universe:New Phenomenology
for the New Millennium?}, hep-ph/9912552.}

\lref\NK{ T. Nihei, {\it Inflation in the five-dimensional universe with an
orbifold extra dimension}, hep-ph/9905487, Phys. Lett. {\bf B465} (1999) 81;
\hfill\break

N. Kaloper, {\it Bent domain walls as brane-worlds}, hep-th/9905210, Phys. 
Rev. {\bf D60} (1999) 123506.}

\lref\Berkl{ C. Cs\'aki, M. Graesser, C. Kolda and J. Terning, {\it Cosmology
of one extra dimension with localized gravity}, hep-ph/9906513, Phys. Lett.
{\bf B462} (1999) 34, hep-ph/9906513.}

\lref\Cline{ J.M. Cline, C. Grojean and G. Servant, {\it Inflating
intersecting branes and remarks on the hierarchy problem}, hep-ph/9909496, 
Phys. Rev. Lett. {\bf 83} (1999) 4245.
}

\lref\Kor{ H.B. Kim and H.D. Kim, {\it Inflation and gauge hierarchy in
Randall-Sundrum compactification}, hep-th/9909053.
}

\lref\moreref{
C. Grojean, J. Cline and G. Servant, {\it Supergravity inspired warped
compactifications and effective cosmological constants}, hep-th/9910081;
\hfill\break  

P. Kraus, {\it Dynamics of anti-de Sitter domain walls}, hep-th/9910149;  
\hfill\break

A. Kehagias and  E. Kiritsis, {\it Mirage cosmology}, hep-th/9910174, JHEP 
9911 (1999) 022; \hfill\break

T. Shiromizu, K. Maeda and M. Sasaki, {\it The Einstein equations on the
3-brane world}, gr-qc/9910076;\hfill\break 

E. Flanagan, S.H.H. Tye and I. Wasserman, {\it Cosmological expansion in the
Randall-Sundrum brane world scenario}, hep-ph/9910498;\hfill\break 

D. Ida, {\it Brane World Cosmology}, gr-qc/9912002.
}

\lref\multibrane{H. Hatanaka, M. Sakamoto, M. Tachibana and K. Takenaga,
{\it Many brane extension of the Randall-Sundrum solution}, 
hep-th/9909076; \hfill\break  

I. Oda, {\it Mass hierarchy and trapping of gravity}, hep-th/9909048;
{\it Mass hierarchy from many domain walls}, hep-th/9908104;} 

\lref\intbrane{N. Kaloper, {\it Crystal manyfold universes in AdS space},
hep-th/9912125; \hfill\break  

T.-J. Li, {\it Noncompact AdS$_5$ universe with parallel positive tension
3-branes}, hep-th/9911234. }

\lref\mop{S. Mouslopoulos and A. Papazoglou, `$+-+$' {\it brane model 
phenomenology}, hep-ph/0003207. }

\lref\sich{S. Ichinose, {\it An exact solution of the Randall-Sundrum model and 
the mass hierarchy problem}, hep-th/0003275.}

\lref\hver{H. Verlinde, {\it Supersymmetry at large distance scales}, 
hep-th/0004003.} 

\lref\prz{L. Pilo, R. Rattazzi and A. Zaffaroni, {\it The fate of the radion in 
models with metastable graviton}, hep-th/0004028.} 

\lref\mps{R.N. Mohapatra, A. Perez-Lorenzana, and C.A. de Sousa Pires, 
{\it Cosmology of brane - bulk models in five-dimensions}, hep-ph/0003328.}

\lref\amo{N. Alonso-Alberca, P. Meessen and T. Ortin, {\it Supersymmetric brane 
worlds}, hep-th/0003248 .}

\lref\abn{R. Altendorfer, J. Bagger and D. Nemeschansky, {\it Supersymmetric 
Randall-Sundrum scenario}, hep-th/0003117.}

\lref\kg{I. I. Kogan, G. G. Ross, {\it Brane universe and multigravity: 
modification of gravity at large and small distances}, hep-th/0003074.}

\lref\kop{P. Kanti, K. A. Olive and M. Pospelov, {\it Static solutions for 
brane models with a bulk scalar field}, hep-ph/0002229.}

\lref\hbk{H. B. Kim, {\it Cosmology of Randall-Sundrum models with an extra 
dimension stabilized by balancing bulk matter}, hep-th/0001209.}

\lref\gac{G. Amelino-Camelia, {\it On a possible quantum limit for the 
stabilization of moduli in brane world scenarios}, hep-th/0001207.}

\lref\dine{M. Dine, {\it Some reflections on moduli, their stabilization and 
cosmology}, hep-th/0001157.}

\lref\rs{L. Randall and R. Sundrum, {\it A Large Mass Hierarchy from a Small 
Extra Dimension}, hep-th/9905221.}

\Title{\vbox{\hbox{hep-ph/0004233}\hbox{MRI-PHY/P20000412}}}
{On Stability of the Three 3-brane Model}

\centerline{\bf Debajyoti Choudhury, Dileep P. Jatkar, Uma Mahanta and Subrata 
Sur\footnote{$^*$}{e-mail: debchou,dileep,mahanta,subrata@mri.ernet.in}}
\medskip
\centerline{\it Mehta Research Institute of Mathematics and Mathematical 
Physics}
\centerline{\it Chhatnag Road, Jhusi, Allahabad 211 019, INDIA}

\vskip .2in
We show that the Goldberger-Wise mechanism for the three 3-brane scenario 
proposed by Kogan et al. stabilizes the radion. We find that the system
of 3-branes stabilizes in such a way that the loss in the scale factor is
insignificant. That is, the negative tension brane chooses to stay close to
the visible brane.
\Date{4/2000}

\newsec{Introduction}

In the quest to explore physics beyond the Standard Model (SM),
certain new proposals\refs{\aadd\rs{--}\rstwo}
have held considerable interest of the community.
Amongst these  
is the proposal by Randall and Sundrum\refs{\rs} wherein the SM fields live 
on one of the two 3-branes which are end of the world branes of a five 
dimensional spacetime. This scenario can be embedded in higher dimensional 
spacetime as well. One of the attractive features of this proposal is the 
resolution of the hierarchy problem between the Planck scale and the 
electroweak scale. This is achieved by choosing the geometry of the embedding 
spacetime with an exponential warp factor for the four dimensional spacetime 
which is the worldvolume of the 3-brane.  
This exponential warp factor produces 
a difference in the mass scales between the two end of the world 3-branes. 

There are several variants of the Randall-Sundrum (RS) model involving
multiple branes\refs{\multibrane,\kg}, intersecting brane 
configurations\refs{\intbrane} and supersymmetry\refs{\hver\amo{--}\abn}. 
Many of these models including the RS model have atleast one negative 
tension brane. This is demanded by the requirement of charge neutrality in the 
compact space. For example, in the original RS model, the visible world lives 
on the negative tension brane 
({\it i.e.}, the one with the induced cosmological constant 
$\Lambda_{vis}<0$).

Any proposal for physics beyond the SM has to pass very stringent tests
laid down by the SM itself as well as by the cosmological observations that 
have been made till date. Brane world cosmology has been studied in 
refs.\refs{\NK\Berkl\Cline\Kor\mps\hbk\dine{--}\moreref}.  These papers 
conclude that the rate of expansion of the universe in the RS model is
different from that of the familiar Friedman-Robertson-Walker cosmology in four
dimensions. In particular, in RS models, the Hubble parameter is proportional
to the energy density $\rho$, i.e., $H\sim \rho$. In contrast, in our 
universe, the Hubble parameter seems to behave as $H\sim\sqrt{\rho}$. It is, 
however, possible to circumvent this problem if one realizes that the total 
energy density $\rho$ is the sum of the vacuum energy density, i.e., the brane 
tension and the matter energy density $\rho_m$ which lives on the brane. In 
this case, dependence of the Hubble parameter on the energy density is 
$H\sim\sqrt{\Lambda_{vis}\rho_m}$. This resolution
brings up a new problem though. If the RS model is to solve the hierarchy 
problem, the visible brane needs to have a negative tension. It turns out 
that the Hubble parameter on such a brane could be real only if the matter 
energy density is negative. However, this would lead to antigravity in the 
brane world, a consequence clearly at variance with experimental observations.

Another issue which was not dealt with in the original RS proposal\refs{\rs} 
was that of the stability of end of the world brane model. This was resolved 
by Goldberger and Wise (GW)\refs{\gw}\footnote{$^1$}{For other as well as 
related proposals of stabilization of the radion modulus see 
\refs{\abn,\kop\prz\sich{--}\gac}.}. They 
introduced a bulk scalar field into the model and showed that its 
(minimal) coupling to the bulk gravity 
stabilizes the two end 
of the world 3-branes at a critical distance $r_c$. For all parameters of 
$o(1)$ this can generate a warp factor of the order $10^{15}$ GeV thereby 
relating the Planck Scale to the SM scale. 

Recently, Kogan et al.\refs{\kmprs} proposed a modified 
version of the RS model 
which contains three 3-branes with both the end of the world branes (one at 
$y=0$ and the other at $y=L_2$) being of positive tension and the third 
(moving) brane with negative tension (at $y=L_1$). We will call this the 
three 3-brane model. The 
visible world lives on one of the end of the world branes 
(at $y=L_2$). This construction automatically ensures 
that the cosmological constant on our brane is positive and 
thereby avoids the problem of the Hubble parameter. This model has 
an exponential warp factor between either of the positive 
tension branes and the negative tension brane. The phenomenological 
consequences are rather striking and have been studied in refs.\refs{\kg,\mop}.

Stability of the Kogan et al.\refs{\kmprs} proposal is quite crucial. This 
is because the moving brane has negative tension and we have exponential 
warp factors growing from the negative tension brane to end of the world 
positive tension branes. Gain in the scale due to
the warp factor between the positive tension brane at $y=0$ and the negative
tension brane at $y=L_1$ is lost partially by the time it reaches our universe 
at $y=L_2$. 
It is easy to see that if this construction of three 3-brane system stabilizes 
for $L_1\sim
L_2/2$ then the loss in the scale hierarchy is near total. It is, therefore, 
crucial that $L_1/L_2 \lappeq 1$. 

In the light of this, we take up the issue of stability of the Kogan et
al. model\refs{\kmprs} in this paper. We use the method of Goldberger and 
Wise\refs{\gw} of coupling a bulk scalar field to the brane system. 
We find that the system does stabilize, and remarkably enough, for 
$L_1/L_2\approx 1$. We also find that this result is not very sensitive 
to small changes in the mass $m$ of the bulk scalar field as well as small 
fluctuations of the parameter $k$ appearing in the warp factor as long as
$m \ll k$.

\newsec{Stabilization of the three 3-brane Model}

In this section, we will first briefly review the Kogan et al. 
model\refs{\kmprs} and will then proceed with the question of modulus 
stabilization. As mentioned in the introduction,
the model of Kogan et al. contains three parallel 3-branes located at $L_0=0$,
$L_1$ and $L_2$. The fifth dimension $y$ has orbifold geometry $S^1/Z_2$ and
$L_0$ and $L_2$ are orbifold fixed points. The action for this configuration
is
\eqn\bulkact{
S = \int d^4x\int_{-L_2}^{L_2} dy
\sqrt{G}\{2M^3R-\Lambda_{bulk}\}-\sum_{i}\int_{y=L_i}d^4x V_i \sqrt{-\hat
G^{(i)}} \ ,
}
where $\hat G^{(i)}_{\mu\nu}$ is the induced metric on the branes and $V_i$
are their tensions. The five dimensional metric ansatz is given by
\eqn\metric{
ds^2 = e^{-2\sigma(y)}\eta_{\mu\nu}dx^{\mu}dx^{\nu} - r_c^2dy^2.
}
This metric preserves four dimensional Poincare invariance. The warp factor
$\sigma(y)$ is just a constant conformal scale factor for the induced four 
dimensional metric. Although it is constant on a given 3-brane, 
its numerical value is different on different
3-branes.

For the three 3-brane model, 
the warp factor grows on either side of the negative 
tension brane up to the positive tension branes. The warp factor $\sigma(y)$
which solves the equations of motion obtained from eq.\bulkact\ is
\eqn\tbsig{
\sigma(y) = k r_c(L_1-||y|-L_1|)
}
when the brane tensions are $V_0=-\Lambda/k$, $V_1=\Lambda/k$ and
$V_2=-\Lambda/k$. Due to the warp factor, the physical mass $M$ on the visible
brane situated at $y=L_2$ is related to the naive mass parameter $M_0$ of the 
four-dimensional Minkowski theory by $M = W M_0$, where, 
\eqn\warpf{W=\exp{\left(-kr_c(2L_1-L_2)\right)}
}
is the warp factor between the visible brane and the other end of the world
brane. As mentioned earlier, to solve the hierarchy problem between the Planck 
scale and the SM scale, $W$ should be of $o(10^{-15})$. Notice that in the 
three 3-brane model, if the moving brane located at $y=L_1$ stabilizes close 
to $L_1 = L_2/2$ then $W\sim 1$. In such a situation, we will not generate 
exponential scale factor in this model despite having the visible brane 
stabilized reasonably far from the other end of the world brane. Thus the 
ratio $L_1/L_2$ is as important, if not more, as the absolute 
stability of the three 3-brane system.

In order to study the stability of the three 3-brane model, consider coupling 
a bulk scalar field to the system of three 3-branes. Our technique is a 
straightforward generalization of the GW mechanism\refs{\gw}. Consider the 
action for a bulk scalar field of mass $m$.
\eqn\action{
S_{\rm Bulk}= {1\over 2}\int d^4x \int_{-L_{2}}^{L_{2}} dy
\sqrt{G}(G^{AB}\partial_A\chi\partial_B\chi -m^2\chi^2),
}
where $G_{AB}$ is the five dimensional metric given in eq.\metric\ with
$\sigma(y)$ given in eq.\tbsig . 

Following ref.\refs{\gw}, we impose the condition that the
bulk scalar field satisfies certain boundary conditions at the location of
each of the branes. The boundary potentials are
\eqn\atzero{\eqalign{
S_0 &= \int d^4x \lambda_0(\chi^2-v_0^2)^2 \quad {\rm at } 
\,\, y=0,\cr S_{L_1} &= \int d^4x \lambda_1(\chi^2-v_1^2)^2 \quad {\rm at } 
\,\, y=L_1,\cr S_{L_2} &= \int d^4x \lambda_2(\chi^2-v_2^2)^2 \quad {\rm at } 
\,\, y=L_2.
}}

We consider only those configuration of the bulk scalar which solve the
equations of motion subject to the condition that the boundary 
potentials (at $y=0$, $L_1$ and $L_2$) are minimised. This essentially amounts 
to neglecting dynamics of $\chi$ along the directions tangential to any of the 
3-branes. This assumption is reasonable because we are looking only at the 
stability of the three 3-brane system at the moment and are not studying the
phenomenologcal consequence of possible coupling of the bulk scalar field 
$\chi$ to matter fields living on the branes. It, therefore, suffices to 
concentrate on the equation of motion of $\chi$ only in $y$ direction. 
Equation of motion for $\chi$ along $y$ is
\eqn\yeom{
\partial^2_y\chi -4\sigma'(y)\partial_y\chi - m^2r_c^2\chi =0
}
where $\sigma'(y) = d\sigma(y)/dy$. It is straightforward to find the solution
to this equation and it is given by
\eqn\soln{
\chi(y) = \exp{(2\sigma'(y)y)}[\tilde A \exp{(\sigma'(y)\nu
y)}+\tilde B \exp{(-\sigma'(y)\nu y)}]
}
where, $\nu = \sqrt{4+(mr_c)^2/\sigma^{\prime 2}(y)}$, $(\tilde A,\tilde B)= 
(A,B)$ for $0\leq y\leq L_1$ and $(\tilde A,\tilde B)= (C,D)$ for 
$L_1\leq y\leq L_2$. It is worth mentioning here that the formal solution
\soln\ is a function of $\sigma'(y)$ but within any given range of values of
$y$, e.g. $0\leq y\leq L_1$, $\sigma'(y)$ is independent of $y$. The 
coefficients $A$, $B$, $C$ and $D$ are determined by demanding that $\chi$ 
minimizes the boundary potential. This gives,
\eqn\abcd{\eqalign{
A &= {v_0 -v_1X^{2-\nu}\over 1-X^{-2\nu}},\quad\quad\quad
B = {v_1X^{2-\nu}-v_0X^{-2\nu}\over 1-X^{-2\nu}}\cr 
C &= {v_1 X^{\nu -2}-v_2 Y^{\nu -2}\over X^{2\nu}-Y^{2\nu}},\quad
D = {-v_1 X^{\nu -2}Y^{2\nu}+v_2 X^{2\nu}Y^{\nu -2}\over X^{2\nu}-Y^{2\nu}}
}}
where, for notational convenience, we have made a change of variables from 
$L_1$ and $L_2$ to $X=\exp{(-kr_cL_1)}$ and $Y=\exp{(-kr_cL_2)}$.
Substituting this solution into the action and integrating out $y$ gives a 
four dimensional effective potential for $X$ and $Y$. Writing this
effective potential in terms of $X$ and a new variable $R\equiv Y/X$, we have,
\eqn\effpot{\eqalign{
k^{-1} V(X, R) &= {1\over 1-X^{2\nu}}\left[(\nu+2)(X^{\nu}
v_0 - X^2 v_1)^2+(\nu -2)(v_0-X^{\nu+2}v_1)^2\right]\cr
&+{X^4\over R^4(1-R^{2\nu})}\left[(\nu+2)(R^2v_1-R^{\nu}v_2)^2+(\nu-2)
(v_2-R^{\nu+2}v_1)^2\right].
}}

An important thing to notice at this point is that, for arbitrary positive 
values of $\nu$, the potential \effpot\
grows as $X\rightarrow 1$ or as $R\rightarrow 1$ as long as $v_1\not= v_0$ and
$v_1 \not= v_2$. These two limits correspond to the negative tension brane 
approaching the positive tension brane at $y=0$ and at $y=L_2$ respectively. 
In fact, the potential has the following singularity in these limits: 
$V(X,R)\sim (1-X^{2\nu})^{-1}>0$ as $X\rightarrow 1$ and 
$V(X,R)\sim (1-R^{2\nu})^{-1}>0$ as $R\rightarrow 1$. This implies that the 
negative tension brane experiences repulsive forces exerted on it by both the 
positive tension end of the world branes
and thus the three 3-brane model cannot reduce to the two 3-brane model. 
In other words,
for a fixed value of $L_2$, $L_1$ lies strictly inside the interval 
$(0, L_2)$. 
However, there is an interesting caveat in this and that has to do with the
choice of vev's of the bulk scalar field on the 3-branes. 
If $v_1 = v_0$ and/or 
$v_1 = v_2$, then the leading singularity in the potential is removed and the
subleading terms in the potential are attractive. In other words, for 
$v_1 = v_2$, 
the potential \effpot\ is such that for $R$ close to $1$, $V$ is
attractive and assumes a  finite value for $R=1$. 
In this case, therefore, the two 3-brane limit is a  stable one. 
The situation is similar, though not identical, 
for $v_1 = v_0$ and $X\rightarrow 1$. {}From here onwards we will work with 
the case when $v_0$, $v_1$ and $v_2$ take different numerical values. 
We still need 
to find out whether it is possible to stabilize $L_2$. An equally important 
question relates to the magnitude of $L_2-L_1$. This number is important to
get the correct metric scale factor on the visible brane.

We will first show how the three 3-brane system is stabilized by the bulk 
scalar field. In the process we
will also be able to determine $L_2-L_1$. Change of variables from $X$, $Y$ to 
$X$ and $R$ has simplified the form of the potential \effpot\ to a 
considerable degree, 
yet, minimization remains a complicated task. The first point to 
notice here is 
that the potential is bilinear in $v_i$ and hence the results depend only on 
the ratios
\eqn\vratios{
  r_{0} \equiv \frac{v_{0}}{v_1}, \qquad 
  r_{2} \equiv \frac{v_{2}}{v_1} \ .
}
Furthermore, this choice of variables is such that the Hessian matrix is
automatically diagonal on the extremal locus.
Extremizing $V$ with respect to $R$, we get
\eqn\Rsoln{
r_2^\mp (R) = 
{\frac{\nu\,{R^{2 + \nu}}\, 
            \left[ \left( 2 \mp {\sqrt{{{\nu}^2} - 4}} \right) \,
          \left( {R^{2\,\nu}} - 1 \right)    + 
       \nu\,\left( 1 + {R^{2\,\nu}} \right)  \right] }{2\,
     \left( {{\nu}^2}\,{R^{2\,\nu}} + 
       2\,{{\left( {R^{2\,\nu}} - 1\right) }^2} + 
       \nu\,\left({R^{4\,\nu}} - 1 \right)  \right) }} \ .
}
\fig{As the ratio $r_2$ of the vev of $\chi$ on the visible brane to that on
the negative tension brane approaches 1, location of these two branes
coincides. However, for $R<1$ we have two possible values for $r_2$ as can be
seen in this graph.}{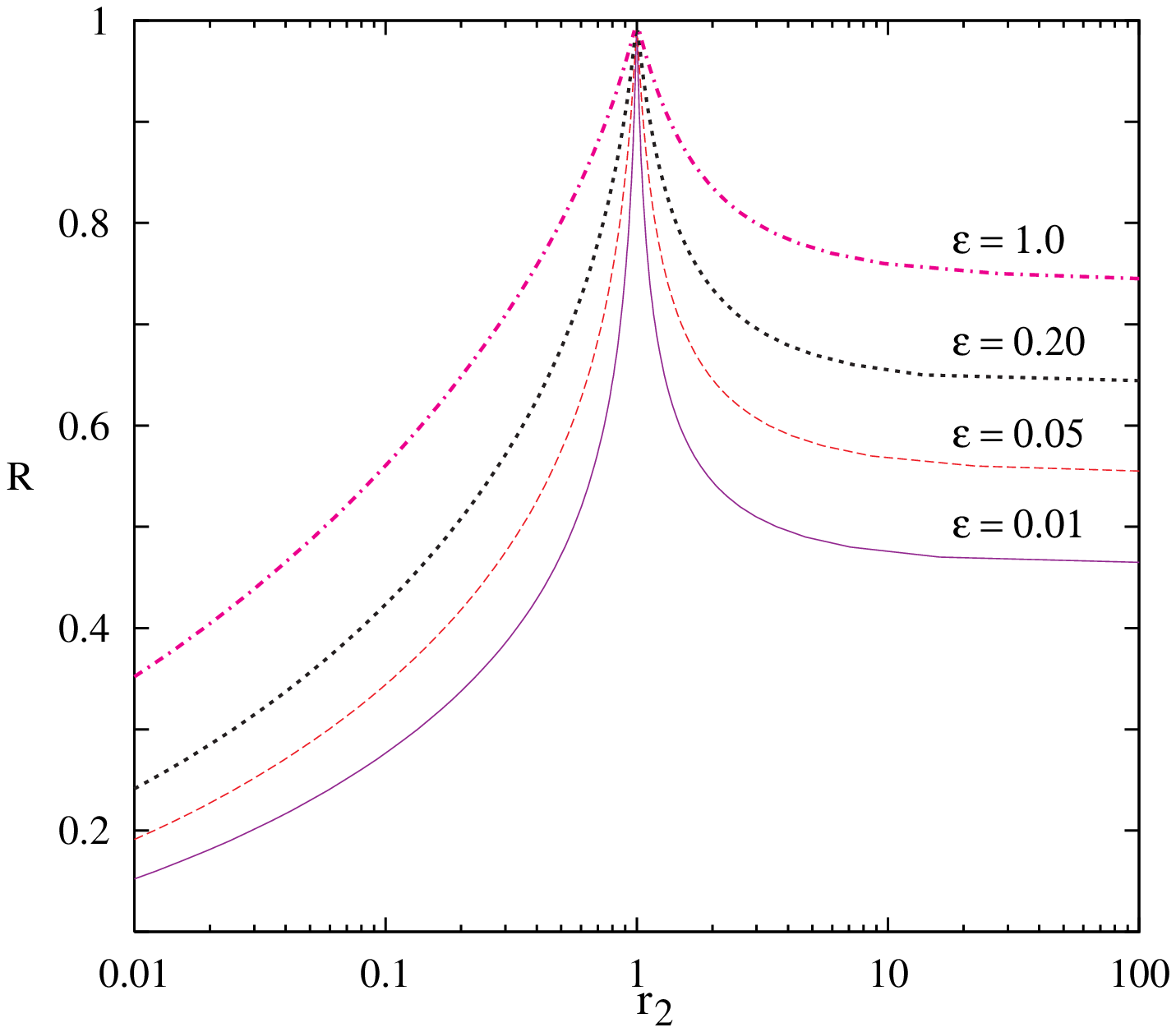}{5truein}\figlabel{\difvev}
An analytic inversion of this equation is not possible on account of its 
transcedental nature.  Numerical solutions are straightforward 
though and we present the results in Fig.\difvev .
For this purpose, instead of working with $\nu$, it is instructive to consider 
\eqn\epsdef{ \epsilon \equiv \nu - 2 \approx \frac{m^2}{4k^2}  \ ,
}
with the approximate equality holding for a light bulk scalar field. 

Although the solutions of eq.\Rsoln\ exist for a wide range of 
$\epsilon$, we are interested in small $\epsilon$'s 
and shall limit ourselves to this regime. In Fig.\difvev, the left and 
right branches correspond to $r_2^-$ and $r_2^+$ respectively. At 
the cusp ($r_2 = 1$), the solutions move into unphysical domains. 
Looking at the figure, we surmise easily that $R \approx 1$ demands 
$r_2 \approx 1$. This is particularly pronounced for small $\epsilon$. 

While the solutions eq.\Rsoln\ represent {\it minima} in the $R$-direction, 
it remains to be seen whether simultaneous minima in the $X$-direction 
exist. We answer this question next. Extremizing with respect to $X$, 
we obtain
\eqn\Xsoln{
r_0^\pm(X; R, r_2) = \frac{X^{2 - \nu}}{2\, \nu} \,
   \left[ 2\, \left( 1 - {X^{2\,\nu}} \right)
	 + {\nu}\, \left( 1 + {X^{2\,\nu}} \right)
\pm 
   \frac{  X^{2 + \nu} ( 1 - X^{2 \nu}) \sqrt{\cal B}  }
        {\sqrt{1 - {R^{2\,\nu}}} }
	\right]   
   \ ,
}
where
\eqn\calB{
{\cal B}  =  
\left( 1 - {R^{2\nu }} \right) \, \left( {{\nu}^2} + 4\nu -12  \right) 
- 8\nu  
+ 16\nu{R^{\nu - 2 }}r_2 
+ 16 {r_2^2}
- 4 {r_2^2} \left(\nu + 2 \right) \left( {R^{2\nu - 4}} + 1\right)
   \ .
}
In each of eq.\Xsoln, $r_2$ assumes both of the values $r_2^\pm$. 
We thus have four possible branches in the solution space. 
Numerically though, for 
given $R$ and $X$, the four solutions for $r_0$ roughly split into two 
pairs (the two $r_0^-$s on the one hand and the two $r_0^+$s on the other), 
with the intra-pair splitting considerably smaller than the inter-pair 
one. More interestingly, for a given $\epsilon$, 
	\item{$\bullet$} not the entire curve of Fig.1 	
	    is admissible when confronted 
	    with eq.\Xsoln. Rather, only $r_2 \approx 1$ (and hence 
	    $R \approx 1$) can lead to a stable solution. For $r_2$ 
	    significantly different from unity, the quantity ${\cal B}$
	    in eq.\calB becomes negative resulting in unphysical values
	    for $X$;
	\item{$\bullet$} for $r_0 = r_0^+$, $\partial^2 V / \partial X^2 < 0 $
	     irrespective of the choice for $r_2$.  Thus, each of these
	    branches correspond to a sequence of saddle points.
%
\par\begingroup\parindent=0pt\leftskip=1cm\rightskip=1cm\parindent=0pt
\baselineskip=11pt
\global\advance\figno by 1
\midinsert 
\vskip -4ex
\centerline{\hskip 3em 
	    \epsfxsize=4.7truein \epsfbox{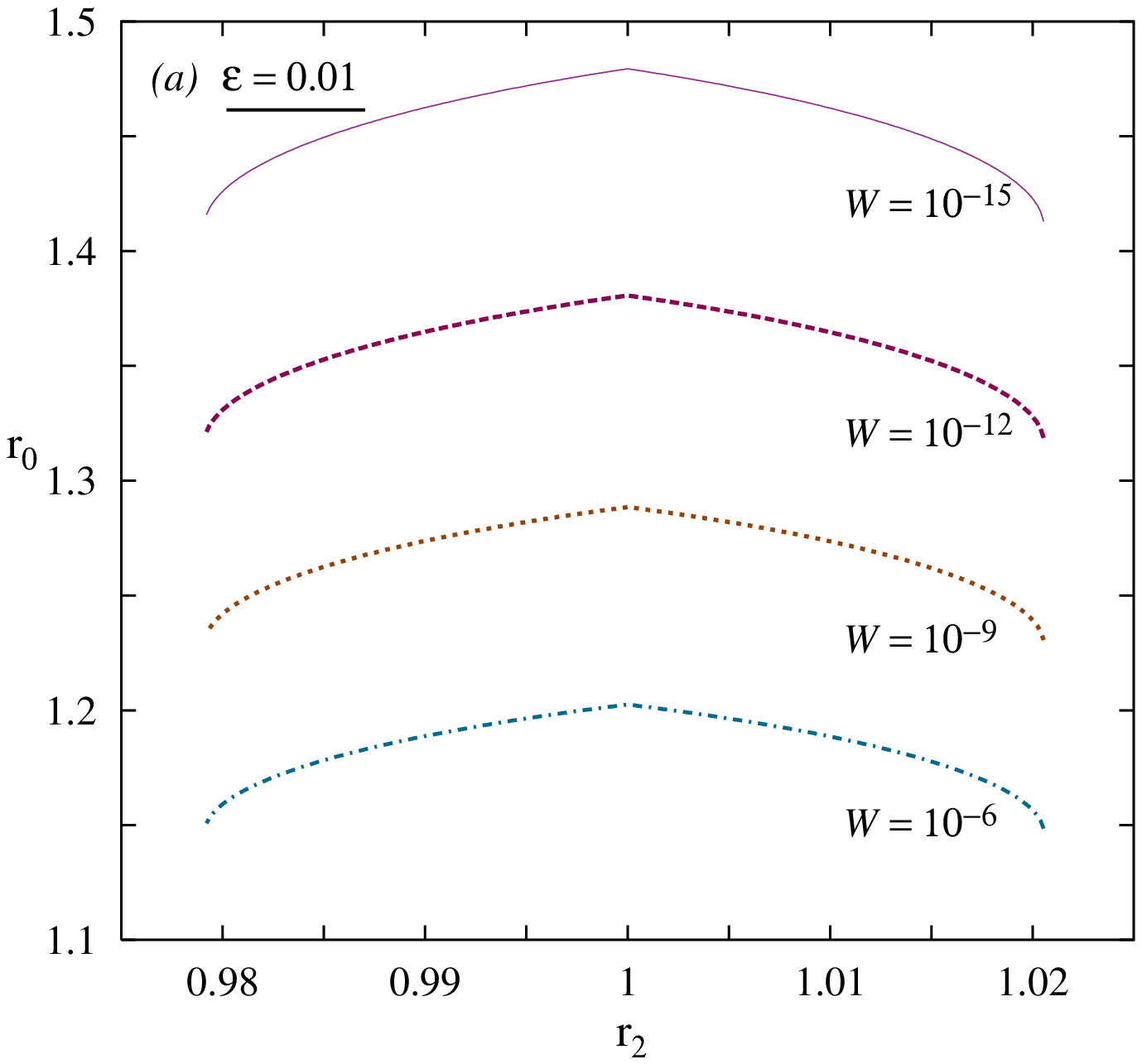} 
	    \hskip -8em
	    \epsfxsize=4.7truein \epsfbox{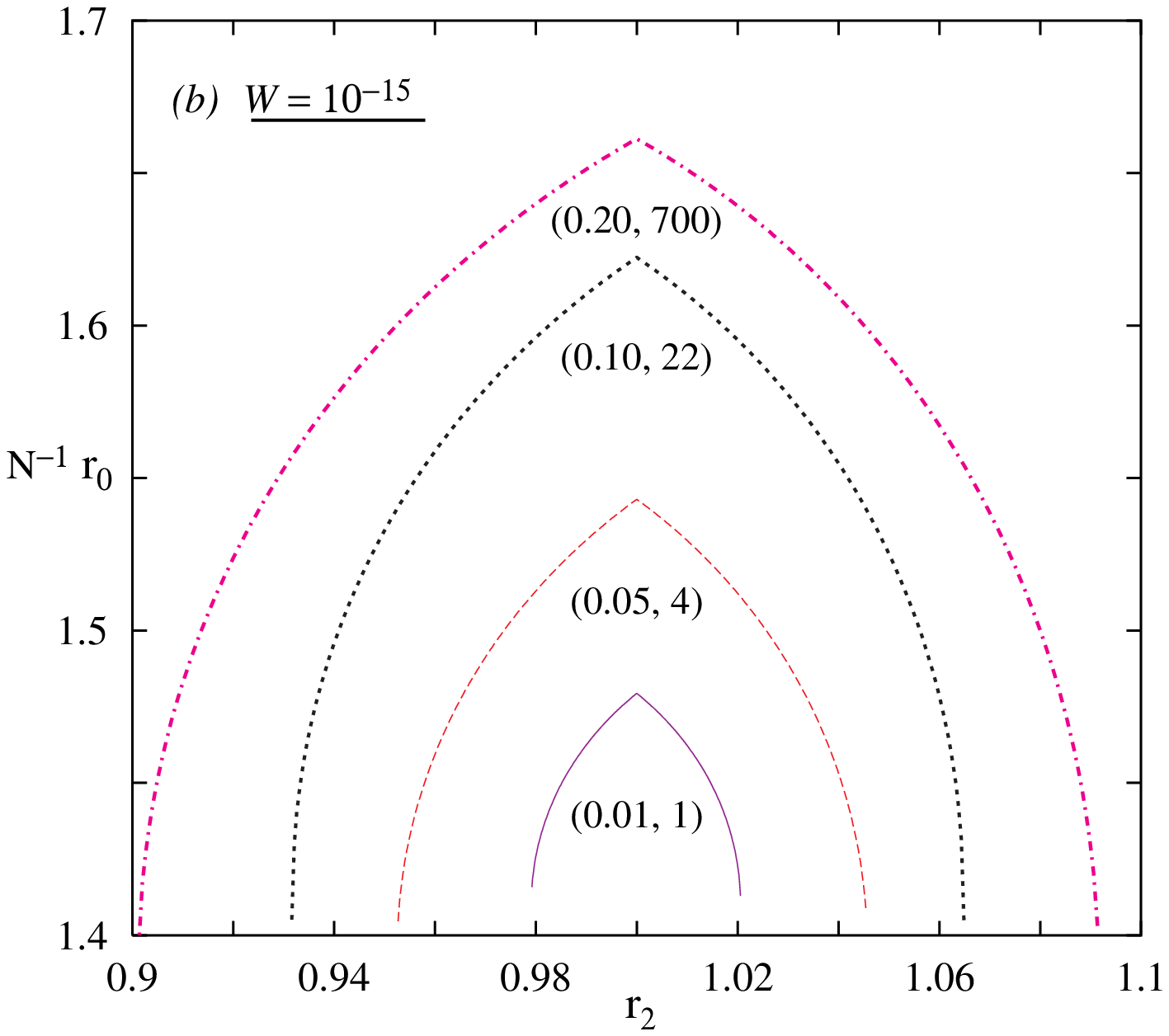}}
\vskip 12pt
{\bf Fig.\ \the\figno:}
   ($a$) For small values of $\epsilon$, the warp factor is not 
	 very sensitive to the vev ratio $r_2$. The dependence 
	 on $r_0$ is rather pronounced though. 
	  Warp factor of $10^{-15}$ can be 
	  obtained with both the ratios $r_0$ and $r_2$ of o(1).
   ($b$) For small $\epsilon$, a warp factor of $10^{-15}$ can be 
	 obtained with both the 
	 ratios $r_0$ and $r_2$ of o(1). However,
	 for larger values of $\epsilon$,  the ratio $r_0$ is 
	 required to be quite large.
	 The numbers in the parentheses refer to ($\epsilon, N$) 
	 where $N$ is the factor by which the $y$-axis has been 
	 rescaled for each curve.
\par
\endinsert\endgroup 
\figlabel{\difscale}
%
\noindent
In Fig.\difscale($a$), we exhibit the relation between $r_0$ and $r_2$ 
that must be satisfied to obtain a particular warp-factor for a given 
$\epsilon$. As is evident, $W$ is a very sensitive function of $r_0$.
This is not very unexpected as, for small $\epsilon$, 
eq.\Xsoln\ essentially gives $r_0$ to be polynomial function 
of  $\ln X$ (and hence $\ln W$). This exponential dependence, a feature 
we share with the GW solution to the 
original RS model, could of course be termed a weakness of such 
stabilization schemes. 

The dependence on $\epsilon$---see Fig.\difscale($b$)---is even more 
pronounced. This, again, is not unexpected as $\epsilon \sim$~o(1)
implies $m \sim k$. For such large masses of a field propagating in 
the anti-de Sitter bulk, its wavefunction decays very fast. 
Consequently, its classical values at the two end of the world branes 
would be widely different. 

\newsec{Discussion}

Brane world universe is one of the promising proposals for exploring the
physics beyond the Standard model. There are several variants of the original
proposal of Randall and Sundrum\refs{\rs}. Many of them are invoked to aviod 
possible contradications with our existing knowledge of the Standard model 
physics and the standard big bang cosmology. The proposal of Kogan et 
al.\refs{\kmprs} is along the same lines. We have shown in this paper that 
the three 3-brane model can be stabilized by coupling the system to a bulk 
scalar field and using the Goldberger-Wise formalism\refs{\gw}.

We find that stability of the three 3-brane model is a more delicate problem
than that for the Randall-Sundrum model. This is because gain in the scale due
to the warp factor in the Randall-Sundrum model is partly offset by the moving
brane. It is therefore not only necessary to have overall stabilization of the
three 3-brane system but it is also necessary to have the ratio $L_1/L_2$ 
close to 1. We show that the generalization of the Goldberger-Wise
mechanism to this model stabilizes the radion modulus in such a way that, for
relatively small mass of the bulk scalar field or, equivalently, small
$\epsilon$, $L_1/L_2 \sim 1$. Consequently, it is possible to generate a scale 
hierarchy between the Planck scale and the TeV scale even without finetuning
the parameters.

As we have discussed earlier, the modulus ceases to be stabilized in the event
that the vevs on two (adjacent) branes are exactly equal. Such a
degeneracy might seem natural, at least in the classical limit. However, it 
should be realized that quantum corrections are likely to disturb such an
equality. This is particularly so since the vev on a brane is affected, at one
loop level and higher, by the particle spectrum on the brane and their
coupling to the bulk scalar field. Unless the spectrum is exactly alike on two
adjacent branes, and unless their coupling to $\chi$ are exactly the same, it
seems unlikely that the two vevs could be equal. And since it is only in this
exactly equal vevs limit, that the potential becomes attractive, the point is,
perhaps, a moot one.
\bigskip

\noindent{\bf Acknowledgments:} We would like to thank A. Sen for discussions.

\listrefs

\bye